\documentclass[usenatbib]{mn2e}

\usepackage{amssymb}
\usepackage{amsmath}
\usepackage{epsfig}
\title{Effects of Biases in Virial Mass Estimation on Cosmic Synchronization of Quasar Accretion}
\author[Charles L. Steinhardt]
       {Charles L. Steinhardt \\
        IPMU, University of Tokyo, 5-1-5 Kashiwanoha, Kashiwa-shi, Chiba-ken, Japan 277-8583}
\date{\today}
\begin{document}

\let\la=\lesssim     
\let\ga=\gtrsim
\def\case#1#2{\hbox{$\frac{#1}{#2}$}}
\def\slantfrac#1#2{\hbox{$\,^#1\!/_#2$}}
\def\onehalf{\slantfrac{1}{2}}
\def\onethird{\slantfrac{1}{3}}
\def\twothirds{\slantfrac{2}{3}}
\def\onequarter{\slantfrac{1}{4}}
\def\threequarters{\slantfrac{3}{4}}
\def\ubvr{\hbox{$U\!BV\!R$}}            
\def\ub{\hbox{$U\!-\!B$}}               
\def\bv{\hbox{$B\!-\!V$}}               
\def\vr{\hbox{$V\!-\!R$}}               
\def\ur{\hbox{$U\!-\!R$}}
\def\ion#1#2{#1$\;${\small\rm\@Roman{#2}}\relax}

\def\aj{\rm{AJ}}                    
\def\apj{\rm{ApJ}}                 
\def\apjl{\rm{ApJ}}                
\def\apjs{\rm{ApJS}}                        
\def\mnras{\rm{MNRAS}}

\maketitle

\label{firstpage}

\begin{abstract}
Recent work using virial mass estimates and the quasar mass-luminosity plane has yielded several new puzzles regarding quasar accretion, including a sub-Eddington boundary on most quasar accretion, near-independence of the accretion rate from properties of the host galaxy, and a cosmic synchronization of accretion among black holes of a common mass.  We consider how these puzzles might change if virial mass estimation turns out to have a systematic bias.  As examples, we consider two recent claims of mass-dependent biases in Mg{\small II} masses.  Under any such correction, the surprising cosmic synchronization of quasar accretion rates and independence from the host galaxy remain.  The slope and location of the sub-Eddington boundary are very sensitive to biases in virial mass estimation, and various mass calibrations appear to favor different possible physical explanations for feedback between the central black hole and its environment.  The alternative mass estimators considered do not simply remove puzzling quasar behavior, but rather replace it with new puzzles that may be more difficult to solve than those using current virial mass estimators and the Shen et al. (2008) catalog.
\end{abstract}



\section{Introduction}

Recent work developing the quasar mass-luminosity plane from a combination of large quasar catalogs \citep{DR7} and virial mass estimators \citep{Vestergaard2006, McLure2004} has provided a new picture of Type 1 quasar accretion seemingly inconsistent with existing models \citep{Steinhardt2010a,Steinhardt2010b}, including:

\medskip
\noindent{\bf 1.  Characteristic Accretion Rate at Fixed $M$ and $z$:} At any fixed combination of mass and redshift, Type 1 quasar accretion is apparently only possible within a narrow ($< 1$ dex) luminosity range around a central, characteristic, sub-Eddington accretion rate.  Therefore, ephemeral properties of host galaxies such as star formation rate and recent merger activity appear to have a minimal effect upon the accretion rate of the central supermassive black hole.  This appears to be in sharp contrast with Seyfert galaxies and other active galactic nuclei with low Eddington ratios, for which properties of the host galaxy are more important.

\medskip
\noindent{\bf 2.  Accretion is Time-Dependent:} The characteristic accretion rate is time-dependent at fixed mass, decreasing towards lower redshift.  Even though galaxies of the same stellar mass may virialize at different times and have different morphologies at the same redshift, the accretion rates of their central supermassive black holes are well synchronized.  

\medskip
\noindent{\bf 3.  Sub-Eddington Accretion:} The characteristic accretion rate is near the Eddington luminosity for the lowest-mass quasars at every redshift, but at higher masses falls increasingly short of Eddington.  Quasars at most combinations of mass and redshift are constrained to lie below Eddington, with the highest mass quasars (and most of the mass growth) taking place closer to $L/L_{Edd} \sim 0.1$.

\medskip
There appears to be a universality to quasar accretion implying that only a few, simple physical processes may play important roles.  If correct, these results may be incompatible with our current understanding not just of quasar accretion and turnoff but perhaps also of structure formation.

These analyses were performed using virial masses provided by \citet{Shen2008} in an extension to the Sloan Digital Sky Survey (SDSS) quasar catalog \citep{Schneider2007}.  Since the publication of these masses, there have been several investigations into whether virial masses might systematically mis-measure black hole masses, particularly for Mg{\small II}-based masses.  Perhaps the existing virial mass scaling relationships must be corrected to account for quasar wind, thermal effects, or other non-virial motion in the quasar broad-line region (BLR) \citep{Onken2008,Risaliti2009,Marconi2009}.  Alternatively, at least the statistical uncertainty in virial mass estimation appears to be better than previously believed, and might even be better than that of more difficult reverberation mapping techniques \citep{Steinhardt2010c}.  \citet{Rafiee2010} claim that different methods of determining the velocity of gas in the quasar BLR show evidence of a mass-dependent, systematic error.  They also claim that correcting for this error returns all quasars to lying at their Eddington luminosity. 

These reports of possible biases in virial mass estimation must also lead us to re-examine the conclusions drawn from the quasar mass-luminosity plane.  Virial mass estimation is inherently a difficult proposition, and estimators are calibrated against only 35 low-redshift quasars with reverberation mapping-based mass estimates \citep{Vestergaard2006,Wang2009,Peterson2004}.  It is therefore essential to understand how sensitive each of the conclusions drawn using the \citet{Shen2008} mass catalog are to biases in virial mass estimation.   

In general, potential biases can be divided into four categories:
\begin{enumerate}
\item{{\bf Affine Biases}: Biases that affinely translate virial mass estimate, shifting, stretching, or contracting the distribution in the $\log M - \log L$ plane.   While this may seem a restrictive definition, because the velocity of broad-line gas is not strongly correlated with its radius, most currently-proposed corrections to virial mass relationships are affine.}
\item{{\bf Correctable Biases}: Biases that do not affect virial mass estimates uniformly across a large catalog, but for which each spectrum contains enough information to correct them.  For example, a hypothetical bias might only affect quasars with strong He{\small II} lines, but could be corrected by properly subtracting He{\small II} before fitting the H$\beta$-O{\small III} complex.}
\item{{\bf Uncorrectable Biases}: Biases that affect different objects in different ways and cannot be corrected from the individual spectra, but can be understood well enough to be modeled in a probabilistic way for large catalogs.  For example, a hypothetical bias might arise from a time delay between variability in the quasar continuum and response of the broad-line region, leading to errors when using monochromatic continuum luminosity to estimate the BLR radius in some objects.}
\item{{\bf Byzantine Biases}: Along the lines of \citet{Pease1980}, arbitrary biases maliciously chosen as a worst-case scenario; one might use these to determine what we have learned with absolute certainty.}
\end{enumerate}

In this work, we will consider the effects of affine biases, a category that includes most currently-proposed corrections.
Both \citet{Risaliti2009} and \citet{Rafiee2010} propose systematic, linear corrections for Mg{\small II}-based estimators relating the true black hole mass $M_{BH}$ and the Shen08 mass $M_{Shen}$ as $M_{BH} = A M_{Shen} + B$.  A {\em compressive} correction with $A > 1$ will act to narrow the quasar mass distribution from its appearance in the Shen08 catalog, while an {\em expansive} correction with $A < 1$ will broaden it.  It is instructive to consider the effects of both types of corrections, biasing the Shen08 masses in different directions, on the key puzzles emanating from the $M-L$ plane analysis of the SDSS catalog.  In \S~\ref{sec:Rafiee}, we consider the proposed correction from \citet{Rafiee2010}, a compressive correction with $A < 1$.  In \S~\ref{sec:Risaliti}, the implications of an expansive correction with $A > 1$, from \citet{Risaliti2009}, are discussed.  Finally, \S~\ref{sec:discussion} considers the lessons learned from this analysis.

\section{A Compressive Correction (Rafiee \& Hall 2010)}
\label{sec:Rafiee}

\citet{Rafiee2010} (RH10) consider the possibility that the the full-width half-maximum (FWHM) of quasar broad emission lines might not directly indicate the velocity of gas in the broad-line region, i.e., that motion in the broad line region may not be predominantly virial.  They instead fit $M_{BH} \propto \textrm{FWHM}^{1.27 \pm 0.40}$ using the MLIN-IX\_ERR fitting method of \citet{Kelly2007}.   Although this possibility might mean abandoning the virial assumption used in virial mass estimation \citep{Steinhardtcomm}, we use it as an example of a possible mass-dependent bias that produces a compressive correction, narrowing the mass distribution at fixed redshift.  

Applying the Rafiee \& Hall (RH) formula produces approximately $\log M_{RH} = 0.67 \log M_{Shen} + 3.0$.  Although the RH10 correction itself is only calibrated for Mg{\small II}, since virial mass estimators using different broad emission lines are calibrated against each other, we will consider the effects of this adjustment at all redshifts, not just $0.4 < z < 2.0$ where Mg{\small II} is available.  A compressive correction such as this does not remove all of the puzzling behavior in the quasar mass-luminosity plane, but does recast some puzzles in a different light.  

\subsection{Puzzling Behavior}

Specifically, this adjustment produces the mass-luminosity plane shown in Figure \ref{fig:rhpanels} in 12 redshift bins ranging from $0.2 < z < 4.1$.  These distributions exhibit the following substantial puzzles:

\begin{figure*}
 \epsfxsize=6in\epsfbox{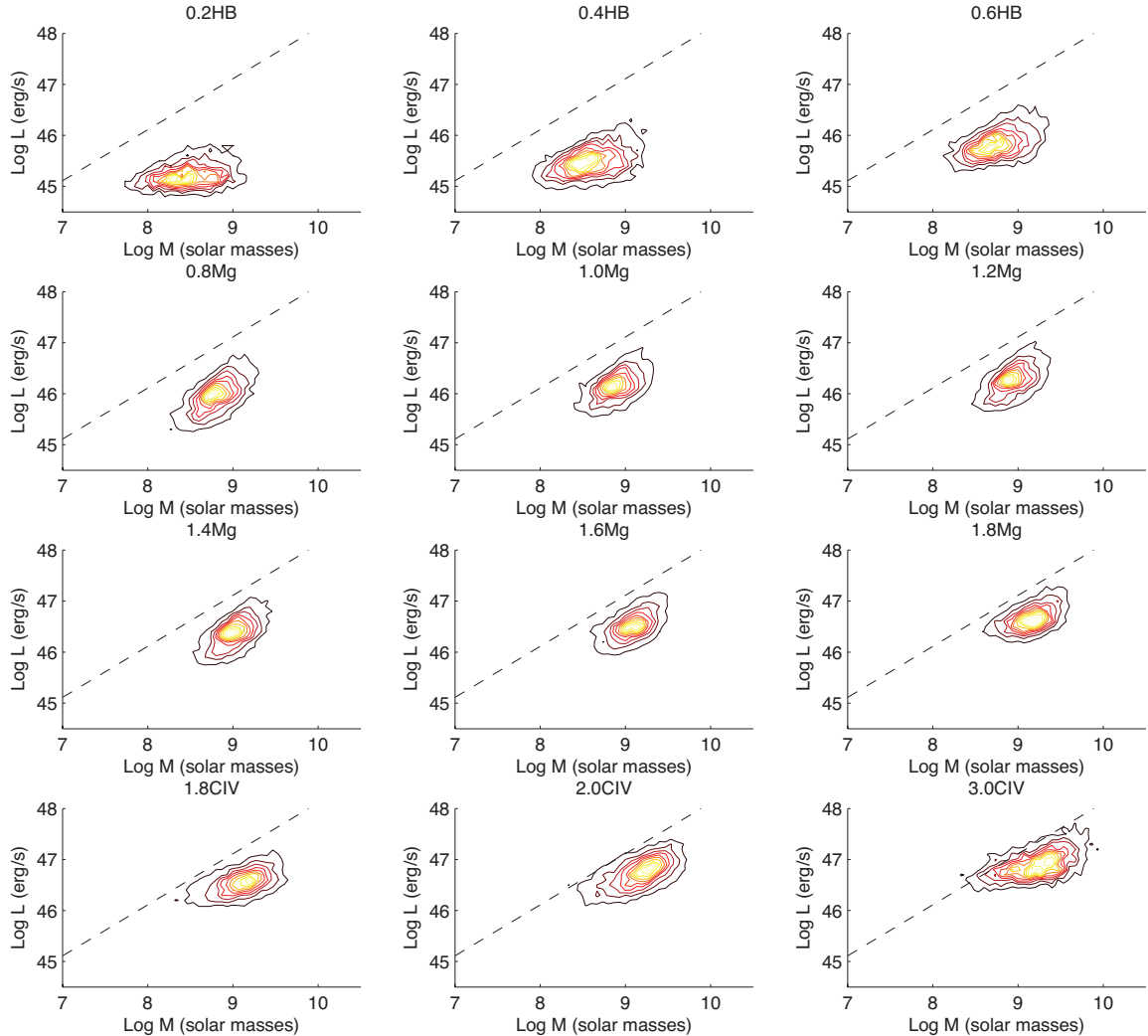}
\caption{A contour plot of the mass-luminosity distribution in 12 different redshift ranges (0.2 - 4.1) using two mass estimators (H$\beta$ and Mg{\small II}) using the Rafiee \& Hall (2010) mass estimators.  Quasars at decreasing redshift lie increasingly below their Eddington luminosity (black, dashed).}
\label{fig:rhpanels}
\end{figure*}

\medskip
\noindent{\bf 1.  Characteristic Accretion Rate at Fixed $M, z$:} At any fixed combination of mass and redshift, the quasar luminosity range spans a range of less than $0.8 - 1$ dex.  Because SDSS views objects across nearly half the sky, the catalog includes quasars that should lie in host galaxies with very different ephemeral properties such as star formation rate and recent merger activity.  Thus, these ephemeral properties can only be important at the $\lesssim 1$ dex level to the quasar accretion rate.  This puzzle also appears when using Shen08 masses.
 
\begin{figure}
 \epsfxsize=3in\epsfbox{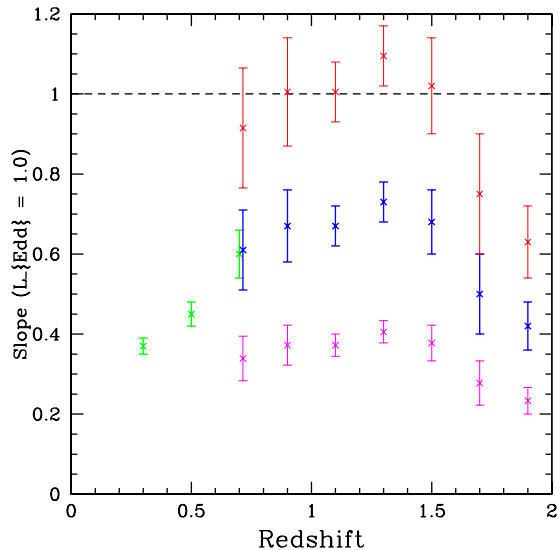}
\caption{Slope of the sub-Eddington boundary as a function of mass for masses estimated using four virial mass calibrators: H$\beta$ (green) and Mg{\small II} (blue) masses from Shen et al. (2008) are compared with Mg{\small II} masses from Rafiee \& Hall (2010) (red) and Risaliti, Young, \& Elvis (2009) (magenta).  Only the Rafiee \& Hall masses yield $L \propto M$ (black, dashed) at any redshift, although no mass estimate is consistent with this dependency at $z \sim 1.9$.}
\label{fig:sebslopes}
\end{figure}

\medskip
\noindent{\bf 2.  Linear Mass Dependence:} At fixed redshift, the characteristic accretion rate increases nearly linearly with mass.  In the Shen08 picture, the characteristic accretion rate increased between $L \propto M_{Shen}^{0.4-0.7}$, and thus for $\log M_{RH} = 0.67 \log M_{Shen} + 3.0$, the characteristic accretion rate will be closer to $L \propto M$.  This is best demonstrated by the behavior of the sub-Eddington boundary, or the luminosity of the most luminous quasars at each mass, using different mass estimation techniques (Fig. \ref{fig:sebslopes}).  The sub-Eddington boundary should run parallel to the characteristic accretion rate, but is easier to measure precisely because it is determined using the brightest and best-measured quasar spectra at each mass and redshift.  Because the slope appears to vary with redshift, the accretion rate cannot be exactly linear in $M$ (or at any other constant value) at all redshifts under any time-independent adjustment to the Shen08 masses. Nevertheless, the accretion rate remains nearly linear.  Since the Eddington luminosity is also $L_{Edd} \propto M$, this means quasars at fixed redshift all lie at nearly the same Eddington ratio $L/L_{Edd}$, rather than at different $L/L_{Edd} \propto M_{Shen}^{-(0.3-0.6)}.$

\medskip
\noindent{\bf 3.  Accretion is Time-Dependent:} The characteristic accretion rate at fixed mass, or chararacteristic Eddington ratio across all masses in the RH10 picture, increases with increasing redshift (Fig. \ref{fig:rhpanels}).  Near $z \sim 2$, quasars are typically able to reach but not exceed their Eddington luminosity.  However, at lower redshift, quasars all miss Eddington by increasingly large margins, and at $z > 3.0$, C{\small IV}-based masses correspond to super-Eddington accretion.  Although C{\small IV} masses have a substantially higher statistical uncertainty than Mg{\small II} \citep{Steinhardt2010c}, suffer from quasar wind and radiation pressure-induced broadening \citep{Marconi2009}, and have a large scatter when compared against Mg{\small II} masses \citep{Shen2008}, there is no systematic disagreement between Mg{\small II} and C{\small IV} masses in the Shen08 sample.  Therefore, corrections inducing a mass-dependent shift in Shen08 Mg{\small II} masses should also apply to C{\small IV} masses, and the resulting statistically significant super-Eddington accretion is either evidence against such a mass correction or yet another puzzle.

\medskip
\noindent{\bf 4.  Narrow Active Mass Range:} At fixed redshift, the observed range in quasar luminosities spans $\sim 1-1.5$ dex, while Shen08 masses span $2-2.5$ dex.  A correction narrowing the mass range to the point that quasars all lie near a fixed Eddington ratio must also result in a mass range spanning $1-1.5$ dex at fixed redshift.  There are three major contributions to the width of this mass distribution:
\begin{itemize}
\item {\bf Mass Growth:} From when they first enter the SDSS catalog, quasars grow in mass via luminous accretion.  Quasars must enter the catalog at least as the lowest-mass objects at the redshift they are observed, and must cease that luminous accretion at a mass no higher than the maximum mass observed {\em at some later redshift}.  Since the maximum mass is decreasing with redshift, the turnoff mass for a quasar in the SDSS catalog must actually be lower than the maximum mass at the redshift it is observed.  The Soltan argument \citep{Soltan1982,Yu2002} demonstrates that most supermassive black hole mass is accreted as a quasar; the mass growth in this phase must be at least $\sim 1$ dex.
\item {\bf Different Timing:}  If all quasars first appeared as the lowest-mass quasars at some redshift, the width of the mass distribution might correspond to the amount of mass growth in a quasar phase.  However, if quasars first appear at higher masses, they will still lie in the observed mass distribution as long as they are quasars, and thus the mass distribution will be wider than the growth of any individual object.  The evolution of quasar host galaxies of a common halo or stellar mass is not well synchronized (cf. \citep{Springel2005}), yet they will eventually produce post-turnoff supermassive black holes of the same mass \citep{msigma1,msigma2}.  The lowest-mass quasars in the SDSS catalog decrease with decreasing redshift due to detector sensitivity.   Thus, if quasars in these host galaxies all turn on at a common mass but at different times, some must not be the lowest-mass quasars when they enter the SDSS catalog.  Differences in the timing of quasar phases will thus lead to a mass distribution broader than one due to mass growth alone.
\item {\bf Mass Errors:} Uncertainties in quasar masses will artificially broaden the quasar distribution; the observed mass distribution is the physical mass distribution convolved with the mass error function.  Statistical errors alone in virial masses have been estimated at $\sim 0.4$ dex \citep{Vestergaard2006}. 
\end{itemize}

For the Soltan argument to apply, luminous accretion must comprise the last $\gtrsim 1$ dex of mass growth.  Thus, the timing of quasar phases in different host galaxies of the same halo mass must be very well synchronized and mass uncertainty must have substantially smaller than an $\sim 0.4$ dex scatter in order to produce a mass distribution spanning $1-1.5$ dex.  Each of these is a puzzle in its own right.

\begin{figure}
 \epsfxsize=3in\epsfbox{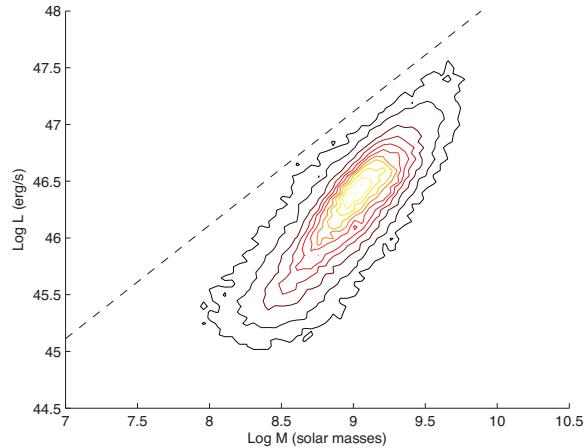}
\caption{The quasar mass-luminosity distribution for all quasars at $0.4 < z < 2.0$, using Mg{\small II} masses as described by Rafiee \& Hall (2010).  Quasars at $\log M/M_\odot  < 8.5$ lie below Eddington, with the shortfall increasing towards lower mass.}
\label{fig:allmg}
\end{figure}

\medskip
\noindent{\bf 5.  Some Masses Never Reach Eddington:} In the Shen08 catalog, with \citet{Vestergaard2006} and \citet{McLure2004} mass estimators, at every redshift there is some mass which quasars reach but do not exceed their Eddington luminosity, and at every mass there is some redshift at which quasars lie at Eddington.  With the Rafiee \& Hall (2010) compressive correction, quasars below $z \sim 1.7$ do not reach Eddington.  Since only a narrow mass range is active at each redshift, quasars with lower masses are never seen at Eddington (Fig. \ref{fig:allmg}).  If, as we are assuming in this section, this is not a calibration problem, then it is a new puzzle requiring a physical explanation.

\subsection{Physical Interpretation}
Like previous mass estimators, the compressive correction proposed by \citet{Rafiee2010} produces a quasar mass-luminosity plane exhibiting several puzzling behaviors.  However, some of these features are new to this calibration, particularly the near-linear dependence of the characteristic accretion rate on black hole mass.  It is natural to believe that the characteristic accretion rate is due to feedback between the central black hole and its environment.  A linear dependence of accretion rate on black hole mass is typical of a model in which gravity is dominant.  Further, x-ray binaries, another type of accretion disk system, are capable of radiatively-efficient accretion only in a Eddington ratio range spanning $\sim 1$ dex; accretion is radiatively inefficient both above and below that range.  Using these mass estimators, quasars, too, can only accrete in an Eddington ratio range spanning $\sim 1$ dex, although the Soltan argument precludes radiatively inefficient accretion at a higher Eddington ratio being common.

If the mass-dependence appears easy to understand, though, the redshift-dependence is quite difficult.  If the accretion rate is set purely by gravity, and feedback from the host galaxy is not important, novel high-energy astrophysics or physics will be needed to explain this time-dependence.  

Another puzzling feature using Shen08 masses is the synchronization of quasar accretion in host galaxies of the same halo mass.  A compressive correction appears to require even stronger, near-perfect synchronization in order to remain consistent with the Soltan argument.  This picture requires either very well synchronization of the first epoch of star formation in each of these host galaxies or, more likely, some sort of primordial black hole seeding mechanism.

\section{An Expansive Correction (Risaliti, Young, \& Elvis 2009)}
\label{sec:Risaliti}

\citet{Risaliti2009} (RYE09) calibrate Mg{\small II}-based masses against H$\beta$, finding that the Mg{\small II} masses systematically disagree with H$\beta$ masses as $\log M_{H\beta} = 1.8 \log M_{MgII} - 6.8$.  As discussed in \citet{Steinhardt2010c}, this may be due to difficulties in separating the Mg{\small II} line and continuum from nearby Fe lines, and as such, the correction may be substantially smaller or unnecessary for the brightest quasars with the highest signal-to-noise spectra at each redshift.  Evidence for this includes good agreement in the sub-Eddington boundary slope calculated using unadjusted H$\beta$ and Mg{\small II} masses at a common redshift, but poor agreement between H$\beta$ and adjusted Mg{\small II} masses (Fig. \ref{fig:sebslopes}).  The best agreement in sub-Eddington boundary slopes comes from a substantially smaller, yet also expansive, correction to Mg{\small II} mass estimates \citep{Steinhardt2010c}.  We will use the RYE09 calibration as an example of a possible mass-dependent bias that produces an expansive correction, widening the mass distribution at fixed redshift.  

\subsection{Puzzling Behavior}

Specifically, this adjustment produces the mass-luminosity plane shown in Figure \ref{fig:ryepanels} in 12 redshift bins ranging from $0.2 < z < 4.1$.  As with the compressive correction considered in \S~\ref{sec:Rafiee}, this does not remove all of the puzzling behavior in the quasar mass-luminosity plane, but rather includes the following surprises:

\begin{figure*}
 \epsfxsize=6in\epsfbox{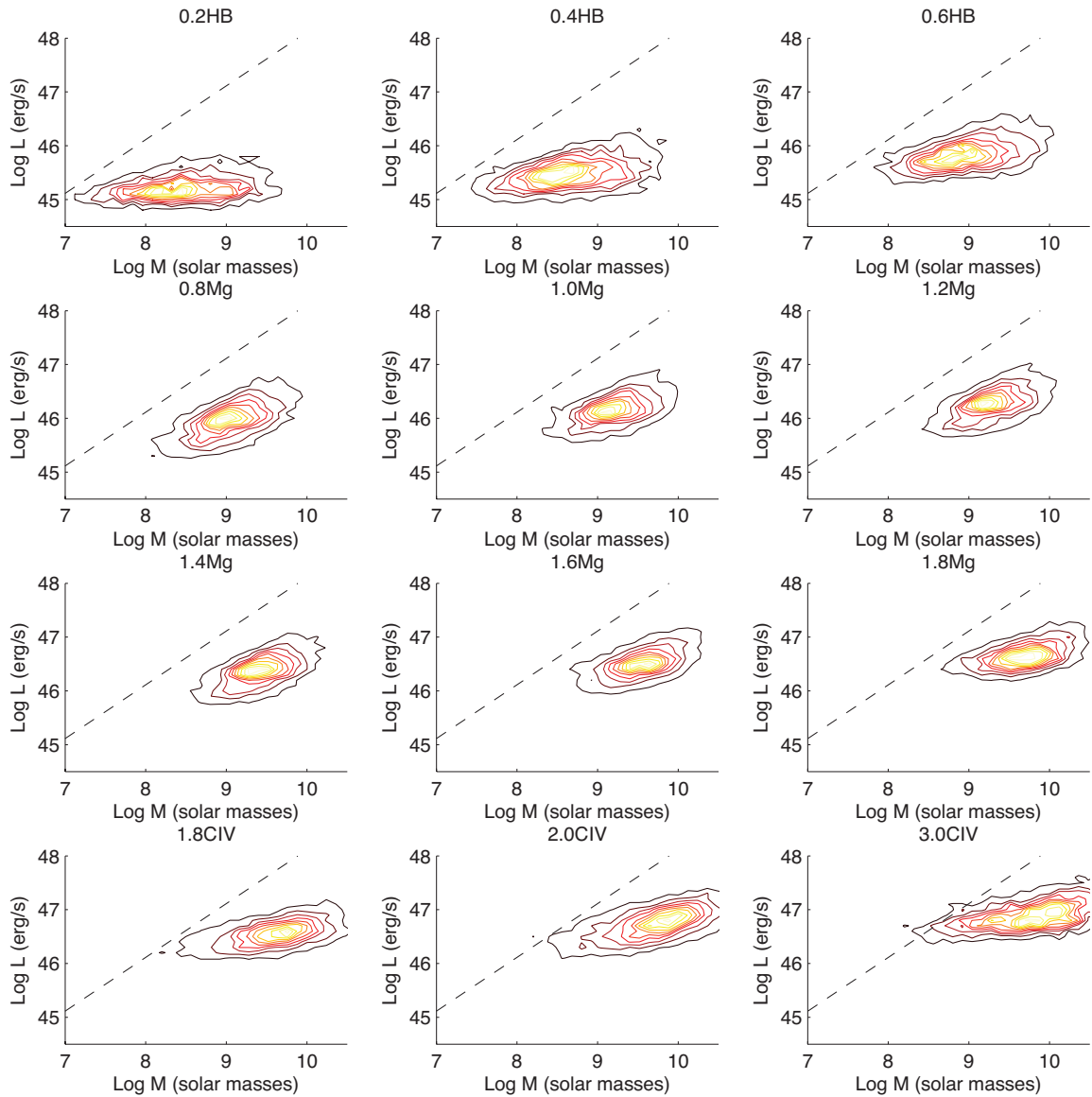}
\caption{A contour plot of the mass-luminosity distribution in 12 different redshift ranges ($0.2 < z < 4.1$) using the Risaliti, Young, \& Elvis mass estimators for Mg{\small II} masses and a corresponding adjustment for H$\beta$ and C{\small IV} masses.  Higher-mass quasars at fixed redshift lie increasingly below their Eddington luminosity (black, dashed).}
\label{fig:ryepanels}
\end{figure*}

\medskip
\noindent{\bf 1.  Characteristic Accretion Rate at Fixed $M, z$:} At any fixed combination of mass and redshift, the quasar luminosity range spans a range of less than $0.8 - 1$ dex.  As before, since SDSS quasars should lie in host galaxies with very different ephemeral properties such as star formation rate and recent merger activity.  Thus, these ephemeral properties can only be important at the $\lesssim 1$ dex level to the quasar accretion rate.  This puzzle also appears when using Shen08 and RH10 masses, and can only be removed by a correction that interprets quasars with different Shen08 masses as having identical black hole masses.

\medskip
\noindent{\bf 2.  Sub-linear Mass Dependence:} At fixed redshift, the characteristic luminosity increases with mass as $L \propto M_{RYE}^{0.2-0.4}$, with the slope closer to $0.4$ at most redshifts (Fig. \ref{fig:sebslopes}).  This dependence is very sensitive to mass-dependent biases, pushing towards a linear mass dependence for extremely compressive corrections and towards only slight mass dependence for extremely expansive corrections.

\medskip
\noindent{\bf 3.  Accretion is Time-Dependent:} The characteristic accretion rate at fixed mass, or chararacteristic Eddington ratio across all masses in the RYE09 picture, increases with increasing redshift (Fig. \ref{fig:ryepanels}).  For example, quasars with $\log M_{RYE}/M_{\odot} = 10^{8.5}$ reach Eddington at $z \sim 2$, but lie increasingly sub-Eddington at lower redshifts.  Even the most luminous quasars at $\log M_{RYE}/M_{\odot} = 10^{8.5}$ and $z \sim 0.2$ lie nearly $1$ dex below their Eddington luminosities.  As with the existence of a characteristic accretion rate, this puzzle also appears using Shen08 and RH10 masses; it can only be removed with a time-dependent mass adjustment.  This may be the least surprising puzzle, as it could be described as quasar downsizing analogous to galactic and star formation \cite{Madau1998} downsizing.

\medskip
\noindent{\bf 4.  Extraordinarily Massive Black Holes:}  The most massive known supermassive black hole, OJ 287, has a mass estimated at $1.8 \times 10^{10} M_\odot$ \citep{Valtonen2008}.  Applying the RYE09 adjustment to Mg{\small II} masses produces a more massive population at $z > 1.6$ (including $\log M_{BH}/M_{\odot} > 10.5$), with the number of more massive objects increasing towards higher redshift.  Although some of these masses use higher-uncertainty C{\small IV}-based estimates, at $1.6 < z < 2.$, these extraordinarily massive black holes are found using Mg{\small II} masses, which are directly calibrated by \citet{Risaliti2009}.  Such immense black holes would need to lie in immense galaxies in order to lie on the $M-\sigma$ relation today.

\medskip
\noindent{\bf 5.  No Eddington Accretion at $z \sim 1$:} At $z \sim 2$ and $z < 0.8$, the least massive quasars reach their Eddington luminosity, while more massive quasars fall increasingly short.  This is an adjusted version of the sub-Eddington boundary described in \citet{Steinhardt2010a}.  However, between these redshifts, all quasars, including those with the lowest black hole masses, fall short of Eddington.  Because the lowest-mass quasars also lie at luminosities near the SDSS detection limit, this may merely be a selection effect.

\subsection{Physical Interpretation}
Like other mass estimators, the expansive correction proposed by \citet{Risaliti2009} produces a quasar mass-luminosity plane exhibiting several puzzling behaviors.  Some of these behaviors, including the narrow luminosity range at fixed mass and redshift and the time-dependence of that characteristic accretion rate, are common to all virial mass estimators.  A unique feature of this correction is an accretion rate $L \propto M_{RYE}^{0.2-0.4}$.  

The accretion rate is likely due to interaction between the central black hole and its environment, and a sub-linear mass-dependence likely indicates a balance between gravitational and non-gravitational effects.  One model with approximately the right mass dependence considers a competition for infalling gas between accretion and star formation \citep{Thompson2005}, yielding $L \propto M^{1/3}$.  

As before, although models may exist for the mass-dependence, the time-dependence is more difficult.  The laws of general relativity are time-independent, while a black hole has no memory; just mass, charge, and spin.  We might expect the feedback mechanism to be sensitive to properties of the host galaxy, but because accretion is well-synchronized between host galaxies containing supermassive black holes of a common mass, ephemeral properties of the galaxy cannot be critical.  As in the Shen08 and RH10 pictures of quasar accretion, the time-dependence continues to defy explanation.  

\section{Discussion}
\label{sec:discussion}

We have considered the effects of affine biases in virial mass estimation by examining two recent reports of biases in different directions \citep{Risaliti2009,Rafiee2010}.  Correcting for one would expand the mass distribution at fixed redshift, while the other would compress it.  Common to all of these virial mass estimators are two major puzzles: (1) that quasars at a common black hole mass and redshift have the same, characteristic luminosity and (2) that the characteristic luminosity is time-dependent.  

It should be immediately apparent that a purely mass-dependent error cannot be responsible for these two results.  Any linear (or more complex) correction to black hole masses in which objects at a common Shen08 mass have a common black hole mass while objects at different Shen08 masses have different black hole masses will yield this same pair of puzzles.  A related puzzle is that using Shen08 masses, the lowest-mass quasars at each redshift are found near their Eddington luminosities, while the highest-mass quasars all fall short of Eddington.  \citet{Rafiee2010} claim to have removed this ``sub-Eddington boundary'' (SEB) with their systematic correction, but because the characteristic luminosity is time-dependent, it is, again, impossible for a purely mass-dependent correction to Mg{\small II} masses to remove this boundary at all choices of redshift.  

Another puzzling feature of the quasar $M-L$ distribution is the mass-dependence of the characteristic luminosity at fixed redshift.  Using the \citet{Shen2008} extension of the SDSS catalog based upon previous mass calibrations \citep{McLure2002,McLure2004,Vestergaard2006} produces $L_{c} \propto M^{0.4-0.7}$, which is presently unmatched by physical modeling.  Unlike the previous two puzzles, the slope of the mass-dependence is highly sensitive to mass correction.  The \citet{Rafiee2010} correction produces a nearly linear dependence, implying that the characteristic accretion rate might be explained due purely to gravity from the central black hole.  Such an explanation would help us to understand why the characteristic accretion rate is universal, and does not depend upon properties of the host galaxy such as star formation rate, merger activity, and age since virialization.  The \citet{Risaliti2009} correction, on the other hand, is a good match for a feedback model describing competition between gravitational attraction and star formation for infalling gas, which again might be independent of ephemeral properties of the host.  

However, neither mechanism can produce all of the the peculiar boundaries in the $M-L$ plane obtained using \citet{Shen2008} masses, most importantly in regard to the puzzling synchronization of quasar accretion.  Moreover, both of these mass corrections introduce new, puzzling features in the $M-L$ plane.  Using the Rafiee \& Hall correction, all quasars will fall short of Eddington at some redshifts, all quasars at some mass will never reach Eddington, and the quasar distribution may be so narrow that the Soltan argument requires perfect synchronization and tiny mass errors. 

Using the Risaliti, Young, \& Elvis correction, again all quasars fall short of Eddington at some redshifts, and many quasars contain black holes more massive than any currently known.  The Shen et al. masses, on the other hand, provide some quasars reaching but not exceeding Eddington at all masses, provide some quasars reaching but not exceeding Eddington at all redshifts, and produce masses lying within the bounds of observed post-turnoff supermassive black holes near redshift zero.  The uncorrected masses appear to be the best of the three options considered, although a smaller correction such as that found in \citet{Steinhardt2010c} by comparing the H$\beta$ and Mg{\small II} views of sub-Eddington boundaries may also be plausible.

\subsection{Further Remarks}
\label{sec:final remarks}

One of the reasons for skepticism about virial masses has been that if taken at face value, the interpretation of the mass-luminosity plane appears inconsistent with what would seem to be obvious properties of supermassive black hole growth.  In particular, supermassive black holes are composed of gas and dust from their host galaxy, live in their host galaxy, and the mass-stellar velocity dispersion ($M-\sigma$) relation \citep{msigma1,msigma2} and black hole mass-galactic luminosity relation \citep{Ferrarese2005} require that SMBH cease their accretion at a final mass dynamically linked to their host galaxy.  It also seems natural that episodes of stronger accretion might be linked to galactic processes such as star formation and mergers that might drive gas and dust into their central regions.

It should be noted that although these puzzles regarding SMBH accretion are new, several other evolutionary phases remain poorly understood.  There are a wide variety of proposed quasar seeding mechanisms \citep{Volonteri2010}, none of which has yet been supported by direct observational evidence.  The black hole mass-stellar velocity dispersion ($M-\sigma$) relation \citep{msigma1,msigma2} and black hole mass-galactic luminosity relation \citep{Ferrarese2005} require that SMBH cease their accretion at a final mass dynamically linked to their host galaxy, but this link is not understood theoretically.  In short, including these new results, we must conclude that we do not understand how SMBH are born, we do not understand how they grow, and we do not understand how they die.  

An inability to understand observational results within the framework of a physical model should indeed be worrying, but in this case, the alternative explanations are no better.  The Sloan Digital Sky Survey quasar catalog appears complete when compared against other quasar catalogs \citep{Jester2005,Steinhardt2010a}, and while the interpretation of virial masses as indicative of the central black hole mass may be in question, our ability to measure the component continuum flux and emission line FWHM is not.  Thus, in order to explain these observations, one of two things must be true:
\begin{itemize}
\item{{\bf Virial masses are not indicative of the central black hole mass.}  Therefore, to explain the observed quasar distribution in $M$, $L$, and $z$, a physical model for accretion is needed in which the BLR velocity is very tightly correlated with the bolometric luminosity, but that correlation is also sharply redshift-dependent.}
\item{{\bf Virial masses do measure the central black hole mass.}  Therefore, to explain the observed quasar distribution in $M$, $L$, and $z$, a physical model is needed in which quasars at fixed mass and redshift truly lie in a narrow range in luminosity, and that narrow range is both mass- and redshift-dependent.}
\end{itemize}

It is unclear which interpretation, if correct, would be more intriguing.  If the BLR velocity is a sufficiently good indicator of bolometric luminosity, a predictive physical model might allow the use of quasars as a bright, high-redshift standard candle.  If virial masses indeed measure central black hole masses, the observed synchroniation of quasar evolution continues until turnoff, at which point the SMBH must lie on the $M-\sigma$ and $M-L$ relations.  Thus, quasars lying in two galaxies of the same mass and same redshift, but with different star formation rates, different morphologies, different merger histories, and different virialization times have the same accretion rate.  So, we are forced to choose between two difficult model-making problems, both requiring a new understanding of quasar accretion, and neither accommodated by existing models.

The author would like to thank Forrest Collman, Martin Elvis, John Silverman, and Michael Strauss for valuable comments.  This work was supported by World Premier International Research Center Initiative (WPI Initiative), MEXT, Japan.

\bibliographystyle{apj}
\bibliography{ms}

\end{document}